\documentclass[
 reprint,
 superscriptaddress,
 amsmath,amssymb,
 aps,
prb,
]{revtex4-2}

\usepackage{graphicx}
\usepackage{dcolumn}
\usepackage{bm}
\usepackage{hyperref}
\usepackage[mathlines]{lineno}
\usepackage{xcolor}
\usepackage{amsmath}
\usepackage{mathtools}
\usepackage{braket}

\begin{document}

\preprint{APS/123-QED}

\title{Crystal electric field excitations and an effective-spin-$1/2$ ground doublet in the hyperkagome magnet Yb$_3$Sc$_2$Ga$_3$O$_{12}$}

\author{Tingjun Zhang}
 \thanks{These authors contributed equally to this work.}
 \affiliation{Department of Physics and Astronomy, Rice University, Houston, TX 77005, USA}
\affiliation{Rice Laboratory for Emergent Magnetic Materials and Smalley-Curl Institute, Rice University, Houston, TX 77005, USA}
\affiliation{Applied Physics Graduate Program, Smalley-Curl Institute, Rice University, Houston, Texas 77005, USA}

\author{Zehao Wang}
 \thanks{These authors contributed equally to this work.}
 \affiliation{Department of Physics and Astronomy, Rice University, Houston, TX 77005, USA}
\affiliation{Rice Laboratory for Emergent Magnetic Materials and Smalley-Curl Institute, Rice University, Houston, TX 77005, USA}

\author{Douglas L. Abernathy}
 \affiliation{Neutron Scattering Division, Oak Ridge National Laboratory, Oak Ridge, Tennessee 37831, USA}

\author{Rong-Zhu Lin}
 \affiliation{Department of Physics and Astronomy, Rice University, Houston, TX 77005, USA}
\affiliation{Rice Laboratory for Emergent Magnetic Materials and Smalley-Curl Institute, Rice University, Houston, TX 77005, USA}
\affiliation{Department of Physics and Center for Quantum Frontiers of Research \& Technology (QFort), National Cheng Kung University, Tainan, Taiwan}

\author{Steven J. Gomez Alvarado}
 \affiliation{Department of Physics and Astronomy, Rice University, Houston, TX 77005, USA}
\affiliation{Rice Laboratory for Emergent Magnetic Materials and Smalley-Curl Institute, Rice University, Houston, TX 77005, USA}
\author{Chien-Lung Huang}
\affiliation{Department of Physics and Center for Quantum Frontiers of Research \& Technology (QFort), National Cheng Kung University, Tainan, Taiwan}
\author{Pengcheng Dai}
\email{pdai@rice.edu}
 \affiliation{Department of Physics and Astronomy, Rice University, Houston, TX 77005, USA}
\affiliation{Rice Laboratory for Emergent Magnetic Materials and Smalley-Curl Institute, Rice University, Houston, TX 77005, USA}

\date{\today}

\begin{abstract}
We use inelastic neutron scattering (INS) to determine the crystal electric field (CEF) excitations of Yb$^{3+}$ in the rare-earth hyperkagome magnet Yb$_3$Sc$_2$Ga$_3$O$_{12}$. Three nearly dispersionless magnetic excitations are observed near 58, 68, and 74~meV, corresponding to transitions from the ground-state Kramers doublet to the three excited doublets of the $J=7/2$ multiplet. A Stevens-operator analysis constrained by the local $222$ ($D_2$) symmetry reproduces the excitation energies and spectral weights and yields an Ising-type ground-state $g$ tensor. The first excited doublet lies approximately 58~meV above the ground state, establishing a well-isolated effective $J_{\mathrm{eff}}=1/2$ degree of freedom over the low-temperature regime relevant to collective magnetism. Notably, the directly measured CEF spectrum substantially revises the level scheme previously inferred from bulk measurements, while preserving the essential low-energy pseudospin description. Two independent fitting protocols give consistent excitation energies, ground-doublet wave functions, and $g$ tensors, despite the nonuniqueness of the individual CEF parameters. The resulting single-ion model also reproduces the characteristic susceptibility, magnetization, and field evolution of the Schottky anomaly in specific heat. These results establish the microscopic single-ion basis needed to construct an effective exchange Hamiltonian and to interpret future measurements of low-energy collective excitations in Yb$_3$Sc$_2$Ga$_3$O$_{12}$.

\end{abstract}

\maketitle


\begin{figure}[t]
\includegraphics[width=\columnwidth]{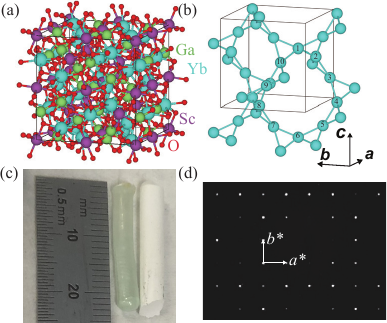} 
\caption{(a) The crystal structure of Yb$_3$Sc$_2$Ga$_3$O$_{12}$. (b) One of the Yb$^{3+}$ hyperkagome networks. A ten-site closed loop is labeled. (c) A representative single crystal grown by the optical floating-zone method (left) and a polycrystalline feed rod used for crystal growth (right). (d) Single crystal X-ray diffraction pattern in the [$H$, $K$, 0] plane.}
\label{fig1}
\end{figure}

\begin{figure*}[t]
\includegraphics[width=\textwidth]{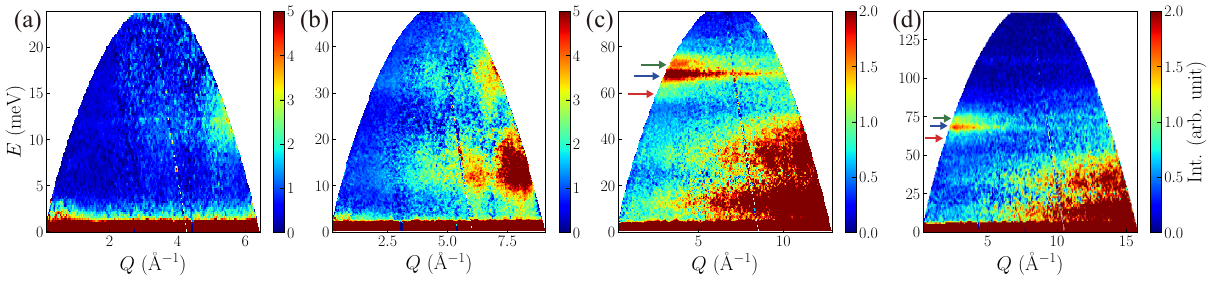} 
\caption{INS data at $T=10$~K measured on powder sample. The incident energies $E_i$ are (a) 25 meV, (b) 50 meV, (c) 100 meV and (d) 150 meV. Empty-can scattering measured at the same temperature has been subtracted. Arrows in (c) and (d) mark the three CEF excitations.}
\label{fig2}
\end{figure*}

\section{Introduction}
In strongly correlated electron systems, single-ion physics plays a central role in determining the low-energy magnetic properties. The single-ion states are first shaped by Coulomb interactions and spin-orbit coupling (SOC), and subsequently by the crystal electric field (CEF) generated by the surrounding ligands. For rare-earth ions in solids, the CEF interaction is typically weaker than the intra-atomic Coulomb interaction and SOC. Coulomb interactions first establish the orbital and spin angular momenta, $\bm{L}$ and $\bm{S}$, which are then coupled by SOC to form the total angular momentum $\bm{J}=\bm{L}+\bm{S}$. The resulting $2J+1$-fold degenerate multiplet is subsequently split by the CEF into a set of local eigenstates. The energies and wave functions of these CEF states determine the single-ion magnetic anisotropy, directional susceptibility, and matrix elements governing magnetic excitations, and thereby strongly influence the collective magnetic behavior \cite{stevensoperator,pointcharge,rareearthmag}. 

The importance of CEF physics has been widely demonstrated in rare-earth magnetic systems. In frustrated magnets, a well-isolated Kramers-doublet ground state can provide an effective pseudospin-$1/2$ degree of freedom, often denoted $J_\text{eff} =1/2$, whose wave function, symmetry, and anisotropic $g$-tensor determine how the pseudospin couples to magnetic fields and to neighboring moments \cite{Gardner2010RMP,Gingras2014RPP,Rau2019ARCM,Smith2025QSI}. Consequently, establishing the CEF level scheme is a prerequisite for constructing microscopic anisotropic exchange models and for assessing whether a material realizes effective $J_\text{eff} =1/2$ magnetism, a key ingredient in many quantum spin ice (QSI) and quantum spin liquid (QSL) candidates. This strategy has been widely applied to three-dimensional (3D) pyrochlore lattices \cite{Gardner2010RMP,Gingras2014RPP,Rau2019ARCM,Smith2025QSI,Gao2019Ce227,gao2025Ce227polarized,gao2026Ce227field111} and two-dimensional (2D) triangular lattices \cite{zhang2021NaYbSe2CEF,pai2022CsYbSe2CEFphonon,li2017YMGOCEF}. Inelastic neutron scattering (INS) is particularly powerful for this purpose because it directly probes both the CEF excitation energies and the associated transition intensities, providing stringent constraints on the CEF Hamiltonian and the corresponding wave functions.

QSLs are exotic magnetic states in which localized spins remain strongly correlated but avoid long-range magnetic order down to the zero temperature limit as a consequence of frustration and quantum fluctuations \cite{zhou2017quantum,savary2017quantum,broholm2020quantum}. The 2D triangular lattice is historically one of the earliest settings proposed to host a QSL state \cite{ANDERSON1973}, where geometrical frustration associated with antiferromagnetic interactions suppresses conventional magnetic order. In rare-earth triangular-lattice materials, strong SOC and the CEF can produce a well-isolated Kramers-doublet ground state described by an effective 
$J_\text{eff} =1/2$ pseudospin with strongly anisotropic magnetic interactions \cite{clark2021qslmaterials}. Together with the reduced dimensionality and geometrical frustration of the triangular lattice, these ingredients have motivated extensive searches for Yb$^{3+}$-based QSL candidates, including YbMgGaO$_4$
\cite{shen2016YMGO,shen2018YMGO,li2016YMGOmuon,luo2018YMGOESR,li2019YMGOreview,zhang2018YMGOTHz,xu2016YMGOthermal,paddison2017YMGOfield,li2015YMGO,li2015YMGOESR} and $A$Yb$Ch_2$ ($A$= Na, K, Rb, Cs; $Ch$= O, S, Se) \cite{baenitz2018NaYbS2,ranjith2019NaYbO2,zhang2021NaYbSe2CEF,zhang2022NaYbSe2musr,dai2021NaYbSe2neutron,ranjith2019NaYbSe2field,scheie2024NaKYbSe2fieldINS,li2024NaYbSe2thermodynamics}. Quantum magnetism can also arise in 3D frustrated lattices, where appropriate lattice topology can generate strong geometrical frustration despite the higher dimensionality. The most prominent examples are the pyrochlore magnets \cite{Gardner2010RMP,Gingras2014RPP,Rau2019ARCM,Smith2025QSI}. One representative QSI candidate is Ce$_2$Zr$_2$O$_7$, in which neutron-scattering studies have established the absence of long-range magnetic order and revealed broad, continuum-like magnetic excitations \cite{Gao2019Ce227,gao2022Ce227fields100110,gao2026Ce227field111,gao2025Ce227polarized,Ce227diffuse,Ce227INS,Ce227poldiff}. Beyond pyrochlores, other 3D frustrated lattices remain considerably less explored. Hyperkagome systems provide one such example, including Na$_4$Ir$_3$O$_8$ \cite{Na4Ir3O8qsl} and Li$_3$Yb$_3$Te$_2$O$_{12}$
 \cite{Li3Yb3Te2O12}. However, microscopic investigations of these materials can be challenging because of practical limitations such as neutron absorption and difficulty in growing sufficiently large single crystals.

\begin{table*}[t]
\caption{CEF coefficients (meV) and principal ground-doublet $g$ factors obtained with the two fitting protocols. The coefficients use the Stevens convention and the local axes defined in Fig.~\ref{fig3}(f).}
\label{Bnm}
\renewcommand{\arraystretch}{1.3}
\begin{ruledtabular}
\begin{tabular}{lcccccccccccc}
Fit & $B_2^0$ & $B_2^2$ & $B_4^0$ & $B_4^2$ & $B_4^4$ & $B_6^0$ & $B_6^2$ & $B_6^4$ & $B_6^6$ & $g_x$ & $g_y$ & $g_z$ \\
\hline
Levels + spectrum & $-0.77$ & $-0.32$ & $-0.016$ & $-0.18$ & $0.12$ & $-8.8\!\times\!10^{-4}$ & $-5.8\!\times\!10^{-3}$ & $2.0\!\times\!10^{-3}$ & $0.012$ & 2.80 & 1.40 & 5.54 \\
Spectrum only & $-0.79$ & $-0.27$ & $-0.015$ & $-0.18$ & $0.11$ & $-8.4\!\times\!10^{-4}$ & $-6.2\!\times\!10^{-3}$ & $1.3\!\times\!10^{-3}$ & $0.012$ & 2.80 & 1.29 & 5.58 \\
\end{tabular}
\end{ruledtabular}
\end{table*}

Recently, Yb$_3$Sc$_2$Ga$_3$O$_{12}$, in which the Yb$^{3+}$ ions form two interpenetrating hyperkagome networks [Figs. 1(a) and 1(b)], has emerged as a promising candidate for realizing a 3D QSL state \cite{Yb3Sc2Ga3O12powder}. Similar to the 2D kagome lattice, each magnetic ion belongs to two corner-sharing triangles and is connected to four nearest-neighbor magnetic ions. In the hyperkagome structure, however, these corner-sharing triangles form an extended 3D network containing characteristic ten-site loops [Fig. 1(b)]. The availability of large single crystals makes Yb$_3$Sc$_2$Ga$_3$O$_{12}$ particularly attractive for comprehensive microscopic studies of 3D quantum magnetism. Measurements of DC and AC magnetic susceptibility, heat capacity, and thermal transport reveal no evidence for long-range magnetic order down to the lowest experimentally accessible temperatures and indicate strong low-temperature spin correlations, consistent with a possible QSL ground state \cite{Yb3Sc2Ga3O12SC}. The magnetization approaches saturation at a relatively modest magnetic field of $B\approx 3$ T \cite{Yb3Sc2Ga3O12SC}, providing an important advantage for INS measurements in the field-polarized state. In this regime, well-defined spin-wave excitations can in principle be measured and quantitatively modeled using linear spin-wave theory, thereby placing strong constraints on the microscopic exchange Hamiltonian. Yb$_3$Sc$_2$Ga$_3$O$_{12}$ therefore provides a rare and experimentally accessible platform for investigating microscopic quantum magnetism on a 3D hyperkagome lattice.


Here, we report INS measurements of the single-ion CEF excitations in Yb$_3$Sc$_2$Ga$_3$O$_{12}$. The three excited Kramers doublets are determined to lie at 58.17, 67.81, and 74.98 meV, establishing a well-isolated $J_{\mathrm{eff}}=1/2$ ground-state doublet separated from the first excited state by an energy scale of approximately 672 K. The resulting single-ion CEF model reproduces the key features of the magnetic susceptibility, magnetization, ESR, and field-dependent heat-capacity measurements \cite{Yb3Sc2Ga3O12powder,Yb3Sc2Ga3O12SC}, thereby establishing the microscopic single-ion basis for future studies of the low-energy collective magnetic excitations.


\section{Methods}
Polycrystalline feed rods were prepared by a conventional solid-state reaction. Stoichiometric amounts of Yb$_2$O$_3$, Sc$_2$O$_3$, and Ga$_2$O$_3$ powders were mixed, ground for 30 min, and sintered at 1100$^{\circ}$C for 24 h. The grinding and sintering procedure was repeated once to improve the phase purity. Single crystals were subsequently grown in a custom-built laser floating-zone furnace\cite{HPFZ, HPFZ_2} using growth conditions adapted from Ref.~\cite{Yb3Sc2Ga3O12SC}. A representative single crystal and feed rod are shown in Fig.~\ref{fig1}(c).

Single-crystal X-ray diffraction measurements were performed using a Rigaku Synergy-S diffractometer with Cu K$_{\alpha}$ radiation ($\lambda \approx 1.54$ \AA). The diffraction data can be indexed within the cubic garnet structure, yielding a lattice parameter of $a=12.38984(13)$ \AA, consistent with previous powder and single-crystal measurements \cite{Yb3Sc2Ga3O12powder,Yb3Sc2Ga3O12SC}. The reconstructed $(H,K,0)$ reciprocal-space plane shown in Fig.~\ref{fig1}(d) further confirms the high single-crystal quality of the grown sample.


For the neutron experiment, approximately 1.5~g of crystal was ground into powder and loaded into an aluminum can. INS measurements were performed on the Wide Angular-Range Chopper Spectrometer (ARCS) at the Spallation Neutron Source, Oak Ridge National Laboratory \cite{abernathy2012ARCS,stone2014SNS}. The sample was cooled to $T=10$~K in a closed-cycle refrigerator. Data were collected with incident energies $E_i=25$, 50, 100, and 150~meV using Fermi-chopper frequencies of 300, 420, 600, and 600~Hz, respectively. Empty-can data measured at 10~K under the corresponding instrumental conditions were subtracted. The data were reduced with \textsc{Mantid} and analyzed with \textsc{DAVE} \cite{mantid,dave}. CEF spectra and single-ion observables were modeled and calculated with \textsc{PyCrystalField} package \cite{pycef}.

\begin{figure}[t]
\includegraphics[width=\columnwidth]{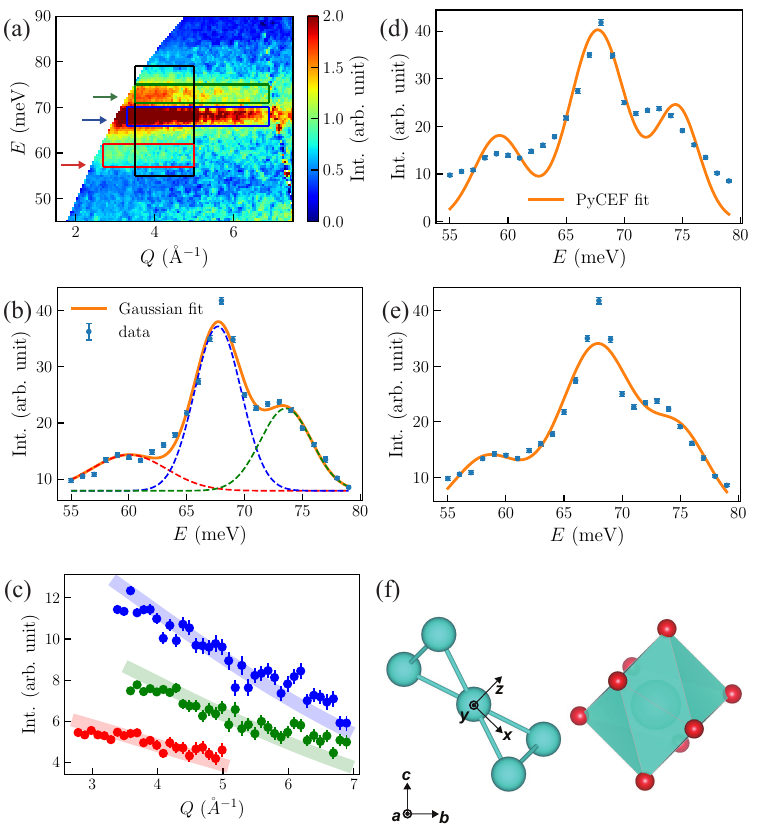} 
\caption{CEF spectrum and modeling. (a) Zoom-in view of the $E_i=100$~meV data. The black rectangle indicates the integration range for the energy cuts in (b), (d), and (e); the colored rectangles indicate the energy windows used for the $Q$ cuts in (c). (b) Energy cut fitted with three Gaussian peaks and a constant background. The weakest peak center was fixed at 59~meV, while the other centers fitted to 67.70(5) and 73.78(9)~meV. (c) Integrated intensity of the three modes versus $Q$; translucent curves show scaled Yb$^{3+}$ magnetic form factor square. (d) CEF fit constrained by both the energy levels and the spectrum. (e) CEF fit to the spectrum alone. (f) Local environment of a Yb$^{3+}$ ion. The local $x$, $y$, and $z$ axes are the principal axes of the $g$ tensor and coincide with the three mutually perpendicular twofold axes of the 222 site symmetry. Left: relative direction of local axes to the hyperkagome lattice. Right: YbO$_8$ polyhedron.}
\label{fig3}
\end{figure}

\section{Results and Discussion}
\subsection{Crystal electric field excitations}

Yb$^{3+}$ has a $4f^{13}$ electronic configuration with $L=3$, $S=1/2$, and a $^2F_{7/2}$ ground-state multiplet. In Yb$_3$Sc$_2$Ga$_3$O$_{12}$, each Yb$^{3+}$ ion is coordinated by eight oxygen ligands, forming a YbO$_8$ polyhedron [Fig.~\ref{fig3}(f)]. Because Yb$^{3+}$ is a Kramers ion, the eightfold-degenerate $J=7/2$ multiplet is split by the CEF generated by the surrounding ligands into four Kramers doublets. At 10~K, only the ground-state doublet is appreciably populated, and therefore up to three CEF transitions from the ground doublet to the three excited doublets are expected.

Figure~\ref{fig2} summarizes the INS spectra. Three sharp, dispersionless excitation modes are clearly resolved in the $E_i=100$ meV data [Fig.~2(c), indicated by the red, blue, and green arrows], and their intensities decrease with increasing momentum transfer $Q$, consistent with a magnetic origin. The same three modes are also observed in the $E_i=150$ meV measurement [Fig.~2(d)], confirming that they are intrinsic excitations rather than instrumental artifacts. Together with the measurements performed using $E_i=25$ and 50 meV [Figs.~2(a) and 2(b)], no additional sharp, dispersionless excitations are observed below approximately 140~meV.


Figure~\ref{fig3}(a) shows the three excitation modes on an expanded energy scale. An energy cut obtained by integrating over the full $Q$ range indicated by the black rectangle is shown in Fig.~\ref{fig3}(b). The spectrum is well described by three Gaussian peaks (red, blue, and green dashed curves) on top of a constant background, consistent with the three CEF transitions expected from the $J=7/2$ ground multiplet. Because the lowest-energy mode is weak and partially overlaps with the tail of a nearby intense peak, its center was fixed at 59~meV during the fit, whereas the other two peak positions refine to 67.70(5) and 73.78(9)~meV. The 59~meV value should therefore be regarded as an approximate experimental peak position rather than an independently refined value. Importantly, two independent CEF-fitting approaches, discussed below, yield closely consistent energies for this lowest excited doublet (Tables~\ref{Eigenvectors} and \ref{Eigenvectors_spec}), supporting the validity of this treatment.


\begin{table*}
\caption{CEF eigenvectors from the levels + spectrum fit. Coefficients are given in the ordered basis $\ket{m_J}=\ket{-7/2},\ket{-5/2},\ldots,\ket{7/2}$. The two rows at each energy form a Kramers doublet.}
\renewcommand{\arraystretch}{1.3}
\begin{ruledtabular}
\begin{tabular}{c|cccccccc}
$E$ (meV) &$| -\frac{7}{2}\rangle$ & $| -\frac{5}{2}\rangle$ & $| -\frac{3}{2}\rangle$ & $| -\frac{1}{2}\rangle$ & $| \frac{1}{2}\rangle$ & $| \frac{3}{2}\rangle$ & $| \frac{5}{2}\rangle$ & $| \frac{7}{2}\rangle$ \tabularnewline
 \hline
0.00 & 0.00 & -0.22 & 0.00 & -0.22 & 0.00 & 0.54 & 0.00 & 0.78 \tabularnewline
0.00 & -0.78 & 0.00 & -0.54 & 0.00 & 0.22 & 0.0 & 0.22 & 0.00 \tabularnewline
59.25 & 0.00 & -0.47 & 0.00 & 0.87 & 0.00 & -0.048 & 0.00 & 0.15 \tabularnewline
59.25 & -0.15 & 0.00 & 0.048 & 0.00 & -0.87 & 0.00 & 0.47 & 0.00 \tabularnewline
67.72 & 0.00 & 0.54 & 0.00 & 0.16 & 0.00 & -0.57 & 0.00 & 0.60 \tabularnewline
67.72 & 0.60 & 0.00 & -0.57 & 0.00 & 0.16 & 0.00 & 0.54 & 0.00 \tabularnewline
74.54 & 0.00 & 0.66 & 0.00 & 0.41 & 0.00 & 0.61 & 0.00 & -0.12 \tabularnewline
74.54 & 0.12 & 0.00 & -0.61 & 0.00 & -0.41 & 0.00 & -0.66 & 0.00 \tabularnewline
\end{tabular}\end{ruledtabular}
\label{Eigenvectors}
\end{table*}

\begin{table*}
\caption{CEF eigenvectors from the spectrum-only fit. Coefficients are given in the ordered basis $\ket{m_J}=\ket{-7/2},\ket{-5/2},\ldots,\ket{7/2}$. The two rows at each energy form a Kramers doublet.}
\renewcommand{\arraystretch}{1.3}
\begin{ruledtabular}
\begin{tabular}{c|cccccccc}
$E$ (meV) &$| -\frac{7}{2}\rangle$ & $| -\frac{5}{2}\rangle$ & $| -\frac{3}{2}\rangle$ & $| -\frac{1}{2}\rangle$ & $| \frac{1}{2}\rangle$ & $| \frac{3}{2}\rangle$ & $| \frac{5}{2}\rangle$ & $| \frac{7}{2}\rangle$ \tabularnewline
 \hline 
0.00 & 0.00 & -0.22 & 0.00 & -0.21 & 0.00 & 0.54 & 0.00 & 0.78 \tabularnewline
0.00 & -0.78 & 0.00 & -0.54 & 0.00 & 0.21 & 0.00 & 0.22 & 0.00 \tabularnewline
58.17 & 0.00 & -0.47 & 0.00 & 0.86 & 0.00 & -0.11 & 0.00 & 0.17 \tabularnewline
58.17 & -0.17 & 0.00 & 0.11 & 0.00 & -0.86 & 0.00 & 0.47 & 0.00 \tabularnewline
67.81 & 0.00 & 0.60 & 0.00 & 0.14 & 0.00 & -0.54 & 0.00 & 0.58 \tabularnewline
67.81 & 0.58 & 0.00 & -0.54 & 0.00 & 0.14 & 0.00 & 0.60 & 0.00 \tabularnewline
74.98 & 0.00 & 0.61 & 0.00 & 0.45 & 0.00 & 0.64 & 0.00 & -0.15 \tabularnewline
74.98 & 0.15 & 0.00 & -0.64 & 0.00 & -0.46 & 0.00 & -0.61 & 0.00 \tabularnewline
\end{tabular}\end{ruledtabular}
\label{Eigenvectors_spec}
\end{table*}

To test the magnetic origin of the observed modes, we integrated the scattering intensity over the energy ranges indicated by the red, blue, and green rectangles in Fig.~\ref{fig3}(a). The integrated intensities of all three modes decrease with increasing momentum transfer $Q$ [Fig.~\ref{fig3}(c)] and are broadly consistent with appropriately scaled Yb$^{3+}$ magnetic form-factor-squared curves, $F^2(Q)$. The measured intensities decrease somewhat more slowly with $Q$ than expected from the magnetic form factor alone, which may reflect residual nonmagnetic background that cannot be fully removed in the absence of a nonmagnetic isostructural reference compound. Taken together, the nearly dispersionless character of the excitations, the observation of the expected number of modes, and their characteristic $Q$ dependence strongly support their assignment as CEF transitions.


\subsection{Crystal electric field Hamiltonian}
We describe the CEF within the $J=7/2$ manifold using the Stevens-operator Hamiltonian

\begin{equation}
H_{\mathrm{CEF}} = \sum_{nm} B_n^m O_n^m,
\end{equation}
where $O_n^m$ are the Stevens operator equivalents, which can be expressed analytically in terms of angular-momentum operators, and $B_n^m$ are the corresponding CEF parameters \cite{CEFhandbook,stevensoperator}. For the local $222$ ($D_2$) symmetry of the Yb site [Fig.~\ref{fig3}(f)], the symmetry-allowed terms are $B_2^0$, $B_2^2$, $B_4^0$, $B_4^2$, $B_4^4$, $B_6^0$, $B_6^2$, $B_6^4$, and $B_6^6$. We adopt the Stevens convention implemented in the \textsc{PyCrystalField} package, with the local coordinate frame defined by the principal axes shown in Fig.~\ref{fig3}(f). Diagonalization of $H_{\mathrm{CEF}}$ within the $^2F_{7/2}$ multiplet of Yb$^{3+}$ yields four Kramers doublets, whose eigenstates can be expressed in the $\lvert J=7/2,m_J\rangle$ basis as:



\begin{equation}
   \ket{\varphi_{\pm}} = \sum_{m_J = -7/2}^{7/2} c_{m_J}^{\pm} \ket{\frac{7}{2}, m_J}.
\end{equation}
where $J=7/2$ is still a good quantum number.

\begin{figure}[t]
\includegraphics[width=\columnwidth]{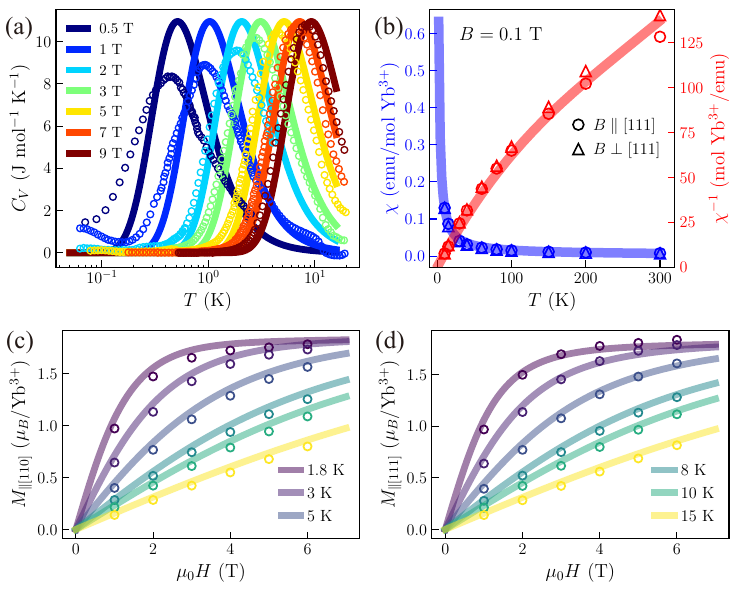} 
\caption{Comparison between single-ion properties calculated from the CEF model and experimentally determined values extracted from Ref.~\cite{Yb3Sc2Ga3O12SC}. (a) Field-induced Schottky heat capacity calculated using $g_{\mathrm{rms}}=3.68$. The phonon contribution to the experimental specific heat was subtracted using the Debye function and coefficients provided in Ref.~\cite{Yb3Sc2Ga3O12SC}. (b) Susceptibility $\chi(T)$ at $\mu_0H=0.1$~T (left axis) and inverse susceptibility (right axis). Because the results for the cubic high-symmetry directions are nearly indistinguishable in the low-field limit, the powder average is used in the calculation. (c),(d) Magnetization for fields along [110] and [111], respectively, at selected temperatures. In Ref. \cite{Yb3Sc2Ga3O12SC}, the authors only present data for fields perpendicular and parallel to [111]. In panel (c), we therefore use the calculated magnetization along [110] as a representative direction perpendicular to [111].}
\label{fig4}
\end{figure}

In determining $H_{\mathrm{CEF}}$, we employed two complementary fitting approaches. In the first approach, the CEF parameters were constrained by simultaneously fitting the three excitation energies obtained from the Gaussian analysis and the measured energy-cut spectrum, including the relative intensities shown in Fig.~3(b). In the second approach, the CEF parameters were determined by fitting the measured spectrum alone \cite{pycef}. The calculated spectra from the two approaches are shown in Figs.~\ref{fig3}(d) and \ref{fig3}(e), respectively, while the fitted CEF parameters and corresponding effective $g$ factors are compared in Table~\ref{Bnm}. The eigenvalues and eigenstates of $H_{\mathrm{CEF}}$ obtained from these parameter sets are listed in Tables~\ref{Eigenvectors} and \ref{Eigenvectors_spec}, respectively.


The two fitting protocols yield CEF excitation energies of $(59.25,67.72,74.54)$~meV and $(58.17,67.81,74.98)$~meV, respectively, together with very similar principal $g$ factors [Table~\ref{Bnm}]. The corresponding CEF eigenvectors are also closely consistent [Tables~\ref{Eigenvectors} and \ref{Eigenvectors_spec}]. Although the calculated line shapes show somewhat larger differences near the individual peak positions, both parameterizations reproduce the overall features of the measured spectrum [Figs.~\ref{fig3}(d) and \ref{fig3}(e)]. Because nine independent CEF parameters $B_n^m$ are being constrained by only three excitation energies and the measured spectral profile, the individual $B_n^m$ values are not expected to be uniquely determined. Nevertheless, the two fitting protocols yield closely consistent excitation energies, relative transition intensities, and representative ground-doublet $g$ tensors. Even the fitted CEF coefficients remain broadly consistent between the two approaches [Table~\ref{Bnm}]. We therefore adopt the spectrum-only parameterization for the analysis below, while regarding it as an effective single-ion CEF model rather than a unique determination of the microscopic CEF coefficients.


\subsection{Ground state doublet anisotropy and single-ion response}

The lowest excited CEF doublet lies approximately 58 meV above the ground state, corresponding to a temperature scale of about 672 K. At temperatures satisfying $k_B T \ll 58$ meV and magnetic fields for which $g\mu_B B \ll 58$ meV, thermal population and field-induced mixing of the excited doublets are negligible. The low-energy magnetic degrees of freedom can therefore be accurately described by a well-isolated $J_{\mathrm{eff}}=1/2$ Kramers doublet, consistent with previous specific-heat measurements and the corresponding magnetic entropy \cite{Yb3Sc2Ga3O12SC,Yb3Sc2Ga3O12powder}.

Projecting the magnetic moment operator onto the ground-state Kramers doublet defines the effective Land\'e $g$ tensor through
\begin{equation}
    \boldsymbol{\mu}=-\mu_B\,\bm{g}\cdot\bm{J}_{\mathrm{eff}},
\end{equation}
where $\bm{J}_{\mathrm{eff}}$ denotes the effective pseudospin-$1/2$ operator. The corresponding principal axes are fixed by the local $222$ symmetry of the YbO$_8$ polyhedron [Fig.~\ref{fig3}(f)]. The local $x$ axis lies along the line common to the two triangles sharing the Yb site and coincides with the angle bisector at that vertex, whereas the local $y$ and $z$ axes are parallel to the sum and difference, respectively, of the normals to the two triangles. Within a chosen Kramers basis $\ket{\varphi_\pm}$, the diagonal expectation values satisfy
\begin{equation}
    \langle \varphi_\pm | J_z | \varphi_\pm \rangle \neq 0,
\end{equation}
while
\begin{equation}
    \langle \varphi_\pm | J_x | \varphi_\pm \rangle
    =
    \langle \varphi_\pm | J_y | \varphi_\pm \rangle
    =
    0.
\end{equation}
The finite transverse $g$ factors instead arise from the off-diagonal matrix elements
\begin{equation}
    \langle \varphi_+ | J_x | \varphi_- \rangle
    \quad \mathrm{and} \quad
    \langle \varphi_+ | J_y | \varphi_- \rangle.
\end{equation}
Because the CEF parameters $B_n^m$ are not uniquely constrained by the INS spectra, the detailed wave functions and individual principal values of the $g$ tensor are also not uniquely determined. Nevertheless, the representative model yields a pronounced hierarchy $g_z>g_x>g_y$ [Table~\ref{Bnm}], revealing a strongly anisotropic ground-state doublet with predominantly Ising-like character along the local $z$ axis.

It is useful to compare this result with the CEF model previously inferred from magnetic and thermodynamic measurements, which placed the three excited Kramers doublets at 69, 96, and 143~meV and yielded principal $g$ factors of $(g_x,g_y,g_z)=(2.93,2.25,4.24)$ \cite{Yb3Sc2Ga3O12powder}. Despite the substantial differences in the excited-state energies, the two models exhibit the same qualitative hierarchy of the principal $g$ factors and comparable orientationally averaged $g$ factors, defined as
\begin{equation}
    g_{\mathrm{rms}}=
    \sqrt{\frac{g_x^2+g_y^2+g_z^2}{3}} \approx 3.68,
\end{equation}
as well as pronounced uniaxial single-ion anisotropy. The direct INS determination, therefore, substantially revises the CEF excitation spectrum while preserving the essential low-energy description in terms of a well-isolated, strongly anisotropic $J_{\mathrm{eff}}=1/2$ Kramers doublet.

Figure~\ref{fig4} shows several single-ion observables calculated from the fitted CEF scheme. For the field-induced Schottky anomaly in magnetic specific heat [Fig.~\ref{fig4}(a)], we use $g_{\mathrm{rms}}=3.68$ together with the experimentally determined zero-field CEF excitation energies. The calculated Schottky peak shifts systematically to higher temperature with increasing magnetic field and shows semiquantitative agreement in both peak position and field evolution with the anomaly observed in heat-capacity measurements \cite{Yb3Sc2Ga3O12powder,Yb3Sc2Ga3O12SC}. At high fields, the calculated specific heat value agrees better with the single-crystal data, whereas a larger discrepancy is observed at lower fields \cite{Yb3Sc2Ga3O12SC}. This discrepancy may indicate that inter-ion interactions make an additional contribution to the low-field heat capacity of the single-crystal sample, beyond the single-ion Schottky contribution captured by the CEF model. Consistent with this interpretation, the powder-sample heat capacity shows much weaker field dependence in the peak magnitude \cite{Yb3Sc2Ga3O12powder}.


The CEF Hamiltonian was also used to calculate the magnetic susceptibility $\chi(T)$ and isothermal magnetization $M(H)$. Owing to the cubic symmetry of space group No.~230 and the presence of 24 symmetry-related Yb$^{3+}$ sites in each unit cell with differently oriented local coordinate frames, the macroscopic magnetic response is expected to be nearly isotropic. Accordingly, the calculated $\chi(T)$ curves for fields applied along the [100], [110], and [111] directions, together with the powder-averaged result, nearly overlap at $\mu_0H=0.1$~T [Fig.~\ref{fig4}(b)], consistent with the weak bulk anisotropy reported for the cubic crystal \cite{Yb3Sc2Ga3O12SC}. The calculated $M(H)$ curves likewise reproduce the principal experimental scales, including a low-temperature moment approaching approximately $1.7$--$1.8~\mu_B$/Yb, near-saturation at fields of order 3~T at 1.8~K, and only weak differences among the high-symmetry field directions [Figs.~\ref{fig4}(c) and \ref{fig4}(d)]. The small discrepancy between the experimental and calculated values may arise from instrumental uncertainties, sample inhomogeneity, and slight deviations of the measurement direction from the exact crystallographic axes. Overall, the consistency across these independent observables provides additional support for the calculated ground-state doublet wave function and effective $g$ tensor, even though the individual CEF parameters are not uniquely determined.
The agreement over an extended temperature range provides a nontrivial consistency check of the CEF eigenstates, since the susceptibility at elevated temperatures also contains contributions from the excited doublets and associated Van Vleck matrix elements.


\section{Conclusion}
We successfully grew high-quality single crystals of Yb$_3$Sc$_2$Ga$_3$O$_{12}$ and identified three CEF excitations near 59, 68, and 74~meV through INS measurements on a powdered crystal sample. A Stevens-operator model consistent with the local $222$ symmetry places the first excited doublet at approximately 58~meV and yields a strongly anisotropic ground-state $g$ tensor. The large CEF gap validates an effective $J_{\mathrm{eff}}=1/2$ description throughout the low-temperature regime relevant to exchange-driven collective magnetism. Two complementary fitting protocols yield robust excitation energies and ground-doublet properties, and the resulting CEF model reproduces the characteristic magnetic-moment scale, saturation field, weak macroscopic anisotropy, ESR $g$ factor, and field evolution of the Schottky anomaly. Our results establish the microscopic single-ion basis needed to construct an exchange Hamiltonian for Yb$_3$Sc$_2$Ga$_3$O$_{12}$ and provide a foundation for future cold-neutron measurements of its low-energy collective excitations.


\begin{acknowledgments}
We thank Sijie Xu and Tong Chen for helpful discussions. The single-crystal synthesis and neutron scattering experiments at Rice were supported by the U.S. DOE, BES under Grant No. DE-SC0026179 (P.D.). Part of the materials characterization efforts at Rice is supported by the Robert A. Welch Foundation Grant No. C-1839 (P.D.). This work was done in part using resources of the Shared Equipment Authority at Rice University (https://research.rice.edu/sea/). We thank Jianhua Li for help with single-crystal X-ray diffraction. R.-Z.L. and C.-L.H. were supported by the National Science and Technology Council in Taiwan (Grant No. NSTC 115-2112-M-006-005. This research was supported in part by an appointment to the ORNL GRO Program, sponsored by the U.S. DOE and administered by the Oak Ridge Institute for Science and Education. This research used resources at the Spallation Neutron Source, DOE Office of Science User Facilities operated by the Oak Ridge National Laboratory (ORNL). ORNL is managed by UT-Battelle, LLC, under contract DE-AC05-00OR22725 with the U.S. DOE. The beam time was allocated to ARCS on proposal number IPTS- 38508.1.
\end{acknowledgments}


\bibliography{YSGO}

\end{document}